\pdfoutput=1
\documentclass[12pt]{article}
\usepackage{graphicx}
\usepackage[T1]{fontenc}
\usepackage{placeins}

\newcommand\pubnumber{DPF2015-1}
\newcommand\pubdate{\today}

\def\cenpa{Center for Experimental Nuclear Physics and Astrophysics, and Department of Physics\\ 
University of Washington, Seattle, WA, USA}
\def\lbnl{Nuclear Science Division, Lawrence Berkeley National Laboratory, Berkeley, CA, USA}
\def\blhill{Department of Physics, Black Hills State University, Spearfish, SD, USA}
\def\ITEP{National Research Center ``Kurchatov Institute'' Institute for Theoretical and Experimental Physics, Moscow, Russia}
\def\JINR{Joint Institute for Nuclear Research, Dubna, Russia}
\def\lbnl{Nuclear Science Division, Lawrence Berkeley National Laboratory, Berkeley, CA, USA}
\def\lanl{Los Alamos National Laboratory, Los Alamos, NM, USA}

\def\ornl{Oak Ridge National Laboratory, Oak Ridge, TN, USA}
\def\ou{Research Center for Nuclear Physics, Osaka University, Ibaraki, Osaka, Japan}
\def\pnnl{Pacific Northwest National Laboratory, Richland, WA, USA}
\def\queens{Department of Physics, Engineering Physics and Astronomy, Queen's University, Kingston, ON, Canada}
\def\ttu{Tennessee Tech University, Cookeville, TN, USA}
\def\sdsmt{South Dakota School of Mines and Technology, Rapid City, SD, USA}
\def\usc{Department of Physics and Astronomy, University of South Carolina, Columbia, SC, USA}
\def\usd{Department of Physics, University of South Dakota, Vermillion, SD, USA} 
\def\ut{Department of Physics and Astronomy, University of Tennessee, Knoxville, TN, USA}

\def\uscornl{Department of Physics and Astronomy, University of South Carolina, Columbia, SC, USA and Oak Ridge National Laboratory, Oak Ridge, TN, USA}
\def\duketunl{Department of Physics, Duke University, Durham, NC, USA and Triangle Universities Nuclear Laboratory, Durham, NC, USA}
\def\unctunl{Department of Physics and Astronomy, University of North Carolina, Chapel Hill, NC, USA and Triangle Universities Nuclear Laboratory, Durham, NC, USA}
\def\ncsutunlornl{Department of Physics, North Carolina State University, Raleigh, NC, USA, Triangle Universities Nuclear Laboratory, Durham, NC, USA, and Oak Ridge National Laboratory, Oak Ridge, TN, USA}
\def\utornl{Department of Physics and Astronomy, University of Tennessee, Knoxville, TN, USA and Oak Ridge National Laboratory, Oak Ridge, TN, USA}
\def\lbnlunctunl{Nuclear Science Division, Lawrence Berkeley National Laboratory, Berkeley, CA, USA, Department of Physics and Astronomy, University of North Carolina, Chapel Hill, NC, USA, and Triangle Universities Nuclear Laboratory, Durham, NC, USA}
\def\unctunlornl{Department of Physics and Astronomy, University of North Carolina, Chapel Hill, NC, USA, Triangle Universities Nuclear Laboratory, Durham, NC, USA, and Oak Ridge National Laboratory, Oak Ridge, TN, USA}

\def\altaddress{\footnote{Alternate address: Department of Nuclear Engineering, University of California, Berkeley, CA, USA}}
\def\Title#1{\begin{center} {\Large #1 } \end{center}}
\def\Author#1{\begin{center}{ \sc #1} \end{center}}
\def\Address#1{\begin{center}{ \it #1} \end{center}}

\newcommand\pubblock{\rightline{\begin{tabular}{l} \pubnumber\\
         \pubdate  \end{tabular}}}
\newenvironment{Abstract}{\begin{quotation}  }{\end{quotation}}
\newenvironment{Presented}{\begin{quotation} \begin{center} 
             PRESENTED AT\end{center}\bigskip 
      \begin{center}\begin{large}}{\end{large}\end{center} \end{quotation}}
\def\Acknowledgments{\bigskip  \bigskip \begin{center} \begin{large}
             \bf ACKNOWLEDGMENTS \end{large}\end{center}}
\def\beq{\begin{equation}}
\def\eeq#1{\label{#1}\end{equation}}
\def\eeqn{\end{equation}}

\def\beqa{\begin{eqnarray}}
\def\eeqa#1{\label{#1}\end{eqnarray}}
\def\eeqan{\end{eqnarray}}

\let\bar=\overbar

\def\Dslash{\not{\hbox{\kern-4pt $D$}}}
\def\dslash{\not{\hbox{\kern-2pt $\del$}}}

\def\msb{{\bar{\ssstyle M \kern -1pt S}}}

\begin{document}
\begin{titlepage}
\pubblock

\vfill
\Title{Status Update of the \textsc{Majorana Demonstrator} Neutrinoless Double Beta Decay Experiment}
\vfill
\Author{J.~Gruszko, M.~Buuck, C.~Cuesta, J.A.~Detwiler, I.S.~Guinn, J.~Leon, and R.G.H.~Robertson}
\Address{\cenpa}
\Author{N.~Abgrall, A.W.~Bradley, Y-D.~Chan, S.~Mertens, and A.W.P.~Poon}
\Address{\lbnl}
\Author{I.J.~Arnquist, E.W.~Hoppe, R.T.~Kouzes, B.D.~LaFerriere, and J.L.~Orrell}
\Address{\pnnl} 
\Author{F.T.~Avignone~III}
\Address{\uscornl}
\Author{A.S.~Barabash, S.I.~Konovalov, and V.~Yumatov}
\Address{\ITEP}
\Author{F.E.~Bertrand, A.~Galindo-Uribarri, D.C.~Radford, R.L.~Varner, B.R.~White, and C.-H.~Yu}
\Address{\ornl}
\Author{V.~Brudanin, M.~Shirchenko, S. Vasilyev, E.~Yakushev, and I.~Zhitnikov}
\Address{\JINR}
\Author{M.~Busch}
\Address{\duketunl}
\Author{D.~Byram, B.R.~Jasinski, and N.~Snyder}
\Address{\usd}
\Author{A.S.~Caldwell, C.D.~Christofferson, C.~Dunagan, S.~Howard, A.M.~Suriano}
\Address{\sdsmt}
\Author{P.-H.~Chu, S.R.~Elliott, J.~Goett, R.~Massarczyk, K.~Rielage, and W.~Xu}
\Address{\lanl}
\Author{Yu.~Efremenko}
\Address{\ut}
\Author{H.~Ejiri}
\Address{\ou}
\Author{T.~Gilliss, G.K.~Giovanetti, R.~Henning, M.A.~Howe, J.~MacMullin, S.J.~Meijer, C. O'Shaughnessy, J.~Rager, B.~Shanks, J.E.~Trimble, and K.~Vorren}
\Address{\unctunl} 
\Author{M.P.~Green}
\Address{\ncsutunlornl}
\Author{V.E.~Guiseppe, D.~Tedeschi, and C. Wiseman}
\Address{\usc}
\Author{K.J.~Keeter}
\Address{\blhill}
\Author{M.F.~Kidd}
\Address{\ttu}
\Author{R.D.~Martin}
\Address{\queens}
\Author{E. Romero-Romero}
\Address{\utornl}
\Author{K.~Vetter\altaddress}
\Address{\lbnlunctunl}
\Author{J.F.~Wilkerson}
\Address{\unctunlornl}

\vfill
\begin{Abstract}
Neutrinoless double beta decay searches play a major role in determining neutrino properties, in particular the Majorana or Dirac nature of the neutrino and the absolute scale of the neutrino mass. The consequences of these searches go beyond neutrino physics, with implications for Grand Unification and leptogenesis. The \textsc{Majorana}  Collaboration is assembling a low-background array of high purity Germanium (HPGe) detectors to search for neutrinoless double-beta decay in $^{76}$Ge. The \textsc{Majorana Demonstrator}, which is currently being constructed and commissioned at the Sanford Underground Research Facility in Lead, South Dakota, will contain 44 kg (30 kg enriched in $^{76}$Ge) of HPGe detectors. Its primary goal is to demonstrate the scalability and background required for a tonne-scale Ge experiment. This is accomplished via a modular design and projected background of less than 3 cnts/tonne-yr in the region of interest. The experiment is currently taking data with the first of its enriched detectors. 
\end{Abstract}
\vfill
\begin{Presented}
DPF 2015\\
The Meeting of the American Physical Society\\
Division of Particles and Fields\\
Ann Arbor, Michigan, August 4--8, 2015\\Majorana
\end{Presented}
\vfill
\end{titlepage}
\def\thefootnote{\fnsymbol{footnote}}
\setcounter{footnote}{0}

\section{Introduction}
In a number of even-even nuclei, $\beta$ decay is energetically forbidden, but the second-order weak process of 2$\nu$ double-$\beta$ decay is allowed. In this rare decay, first proposed by Goeppert-Mayer in 1935 \cite{GoeppertMayer}, $(A, Z) \rightarrow (A, Z+2) + 2 e^{-} +2 \nu$. If the neutrino is a Majorana particle, neutrinoless double-$\beta$ decay could also occur via the exchange of a light Majorana neutrino, or by other mechanisms \cite{Avignone2008}. This decay, $(A, Z) \rightarrow (A, Z+2) + 2\beta$, violates lepton number and provides a model-independent test of the nature of the neutrino. 

\begin{figure}[htb]
\centering
\includegraphics[height=1.5 in]{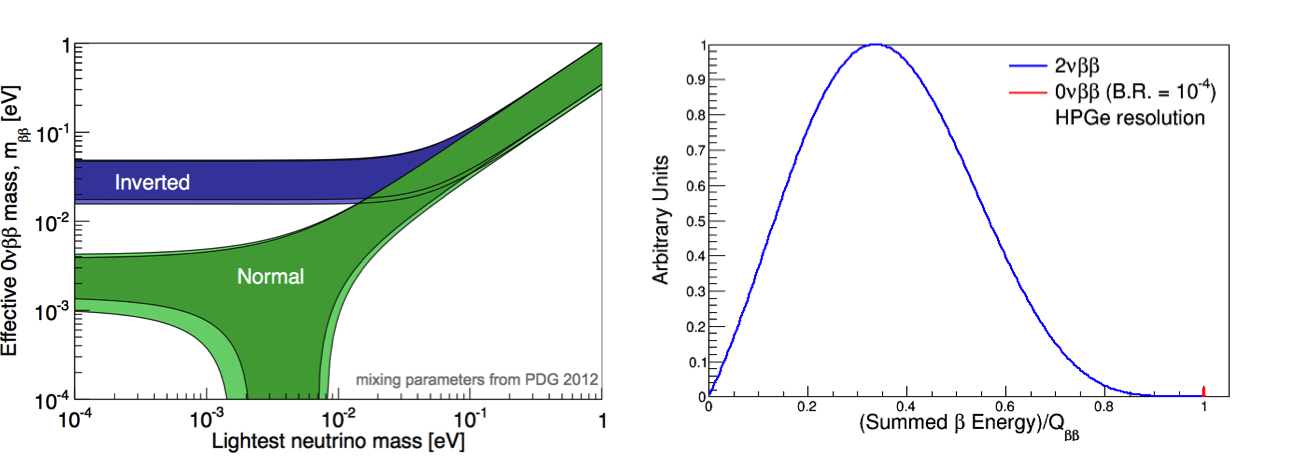}
\caption{\textit{Left:} $\langle m_{\beta\beta} \rangle$, and therefore the $0\nu\beta\beta$ rate, depend on the mass hierarchy and overall mass scale. \cite{SchubertThesis} \textit{Right:} The experimental signature of $0\nu\beta\beta$ decay as it would appear in $^{76}${Ge}.}
\label{fig:BBplots}
\end{figure}

The rate of neutrinoless double-$\beta$ ($0\nu\beta\beta$) decay via light Majorana neutrino exchange is given by 
$$ (T_{1/2}^{0\nu})^{-1} = G^{0\nu}|M_{0\nu}|^{2}\left(\frac{\langle m_{\beta\beta} \rangle}{m_e}\right)^2 $$
where $G^{0\nu}$ is a phase space factor, $M_{0\nu}$ is the nuclear matrix element, and $m_e$ is the electron mass. $\langle m_{\beta\beta} \rangle$ is the effective Majorana mass of the exchanged neutrino, $\langle m_{\beta\beta} \rangle = |\sum\limits_{i=1}^3 U^2_{ei}m_i|,$ where $U_{ei}$ specifies the admixture of neutrino mass eigenstate $i$ in the electron neutrino. Because $\langle m_{\beta\beta} \rangle$ depends on the oscillation parameters, both the overall neutrino mass and the mass hierarchy can contribute to the observed rate (see Fig.~\ref{fig:BBplots}). Provided the nuclear matrix elements are understood, $0\nu\beta\beta$ decay experiments could establish an absolute scale for the neutrino mass. 

The experimental signature of such a decay would be a peak in energy at the endpoint of the two-neutrino mode spectrum with width defined by the experiment's energy resolution, as shown in Fig.~\ref{fig:BBplots}. Given the extremely low rates predicted (current limits indicate $T_{1/2} > 10^{25}$ years) and irreducible background due to the 2$\nu$ mode, high source mass, high efficiency, excellent resolution and extremely low backgrounds in the signal region are key.   

In the previous generation of experiments, the most sensitive limits on $0\nu\beta\beta$ decay came from the IGEX \cite{IGEX_result, IGEX_response} and Heidelberg-Moscow \cite{HeidelbergMoscow_result} experiments, both using $^{76}$Ge. An observation of $0\nu\beta\beta$ decay was claimed by a subgroup of the Heidelberg-Moscow collaboration \cite{KK2001}. Recent searches carried out in $^{76}$Ge (GERDA \cite{GERDA2013}), in $^{136}$Xe (EXO-200 \cite{EXO2014}, KamLAND-ZEN \cite{KAMLAND2012}), and in $^{130}$Te (CUORE-0 \cite{CUORE2015}) have set limits that do not support such a claim. 

\section{The \textsc{Majorana Demonstrator}}
\subsection{Overview}
The \textsc{Majorana Demonstrator} (MJD) \cite{MJD_overview} is an array of enriched and natural germanium detectors that will search for the $0\nu\beta\beta$ decay of $^{76}$Ge. Its main goal is to achieve backgrounds of 3 counts/tonne-year in the 4 keV region of interest (ROI) around the 2039 keV Q$_{\beta\beta}$ of $^{76}$Ge $0\nu\beta\beta$ decay after analysis cuts. This background level, which scales to 1 count/ROI-t-y in a tonne-scale experiment, is the most aggressive background goal of any current experiment. MJD's additional goals are to establish the feasibility of constructing and fielding modular arrays of germanium detectors and to conduct searches for other physics beyond the standard model, such as WIMP dark matter and axions.  

\begin{figure}[htb]
\centering
\includegraphics[width = .65 \textwidth]{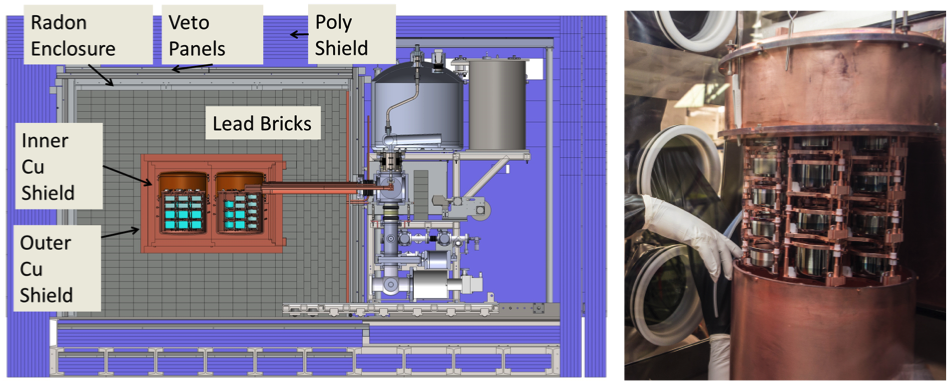}
\caption{\textit{Left:} A model of the \textsc{Majorana}  DEMONSTRATOR, showing the shielding and veto system. \textit{Right:} The cryostat of Module~1, being sealed while inside the glovebox.\protect\footnotemark}
\label{fig:MJD_model}
\end{figure}
\
\footnotetext{Photo from http://pics.sanfordlab.org/} 
MJD consists of 44 kg of P-type point-contact (PPC) germanium detectors. 29.7 kg of this material is enriched to 87\% $^{76}$Ge, contained in 35 detectors provided by ORTEC, and the remaining 15~kg is natural-abundance germanium, contained in 24 BEGe detectors from Canberra. The PPC geometry was chosen for two reasons: its small ($\sim$1 pF) capacitance, which leads to good energy resolution ($<$\,2.3~keV FWHM at 1333~keV) and low energy thresholds ($\mathcal{O}$(a few keV)), and its localized weighting potential, which allows for multi-site event rejection. 

MJD uses a modular design, making it naturally scalable to tonne-scale and allowing for staged deployment. Two ultra-clean electroformed copper cryostats each contain seven ``strings'' of three to five detectors each. The two cryostats have independent vacuum and cryogenic systems, and share a compact passive shield and 4$\pi$ active muon veto system. See Fig.~\ref{fig:MJD_model}. Implementation is divided into three stages: (1) The Prototype Cryostat, used for R\&D, which was made from OFHC copper and contained ten natural Ge detectors, (2) Module~1, which contains 16.8~kg of enriched Ge (20 detectors) and 5.7~kg of natural Ge (9 detectors), and (3) Module~2, which contains 12.6~kg of enriched Ge (14 detectors) and 9.4~kg of natural Ge (15 detectors). 

\subsection{Background Reduction}
MJD is being built and housed at the 4850' level (4260~m.w.e. overburden) of the Sanford Underground Research Facility (SURF) \cite{SURF_overview} in Lead, South Dakota, to reduce cosmogenic activation of materials and muon flux in the experiment. Construction is being done in a class 1000 cleanroom to limit contamination from natural radioactivity, with ultra-low background components (i.e. all parts internal to the cryostats) assembled in a class 10 nitrogen-purged glovebox, to reduce radon contamination. Most low-background parts are machined in an underground shop to control contamination and cosmogenic activation, and tracked extensively \cite{MJD_PTDB}.
\begin{figure}[htb]
\centering
\includegraphics[height=3 in]{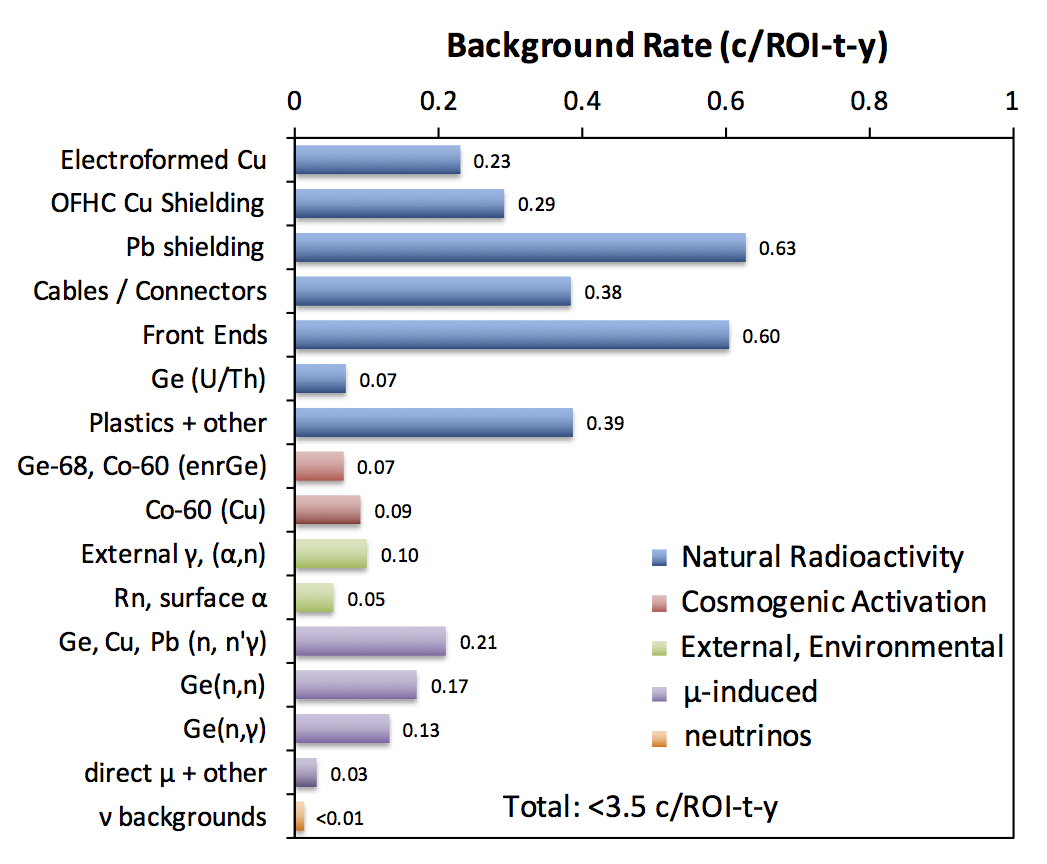}
\caption{A breakdown of expected background contributions in the \textsc{Majorana Demonstrator}.}
\label{fig:bgBudget}
\end{figure}
\
To reduce background contributions from natural radioactivity, the \textsc{Majorana} Collaboration has conducted an extensive assay campaign using gamma counting, NAA, ICP-MS, and GDMS. Most components of the experiment are made from ultra-clean copper that has been electroformed underground at SURF and Pacific Northwest National Laboratories. Current assays indicate that this copper has less than 0.1~$\mu$Bq/kg of activity in each of the uranium and thorium decay chains. All other components are low-mass and use low-background materials. Based on assay values and upper limits, the background rate is projected to be $<$\,3.5~counts/ROI-t-y. See Fig.~\ref{fig:bgBudget} for details. 
\begin{figure}[htb]
\centering
\includegraphics[width = \textwidth]{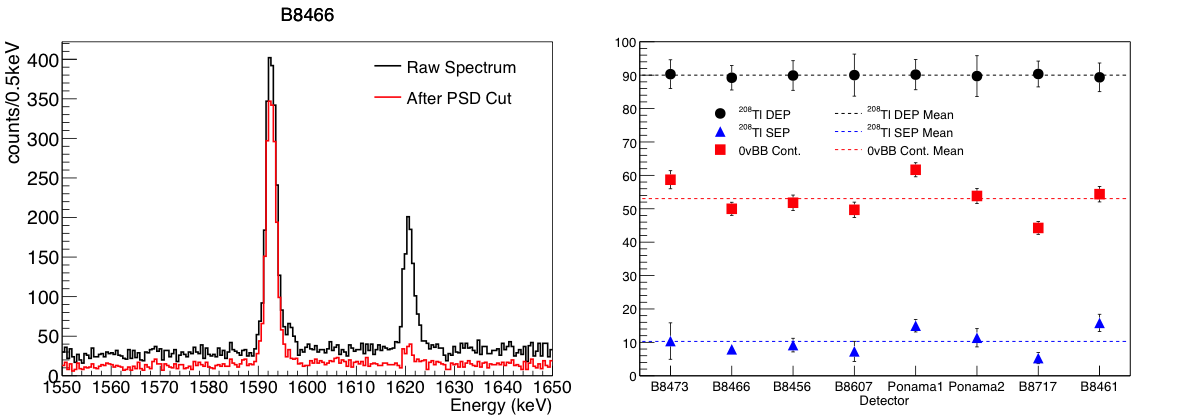}
\caption{\textit{Left:} The energy spectrum for a natural BEGe detector in the Prototype Cryostat shows that multi-site events are cut effectively, while retaining high efficiency for a nearby single-site peak. \textit{Right:} In prototype cryostat calibration runs, pulse shape discrimination methods are tuned to retain 90\% of known single-site events, found in the $^{208}$Th double escape peak. They reject >\,90\% of known multi-site events, found in the $^{208}$Th single escape peak. At Q$_{\beta\beta}$, the Compton continuum is reduced by 50\%.}
\label{fig:PSD}
\end{figure}
\

Powerful background-rejection analysis techniques further enable the \textsc{Demonstrator}'s physics reach. The goal of such techniques is to reject multiple interaction site events due to gamma rays, since they are not plausible double-beta decay candidate events. Using pulse shape discrimination methods \cite{BEGe_PSD}, it is possible reject 90\% of multi-site events while retaining 90\% of single-site events and reducing the Compton continuum at $Q_{\beta\beta}$ by approximately 50\%, as seen in Fig.~\ref{fig:PSD}.   
\section{Status of the \textsc{Majorana Demonstrator}}
The Prototype Module, which included a commercial copper cryostat, was used for research and development of mechanical systems, fabrication and cleaning processes, and assembly procedures. It took data in the shield from July 2014 to June 2015, and has now been decommissioned. 

\begin{figure}[htb]
\centering
\includegraphics[width = .6\textwidth]{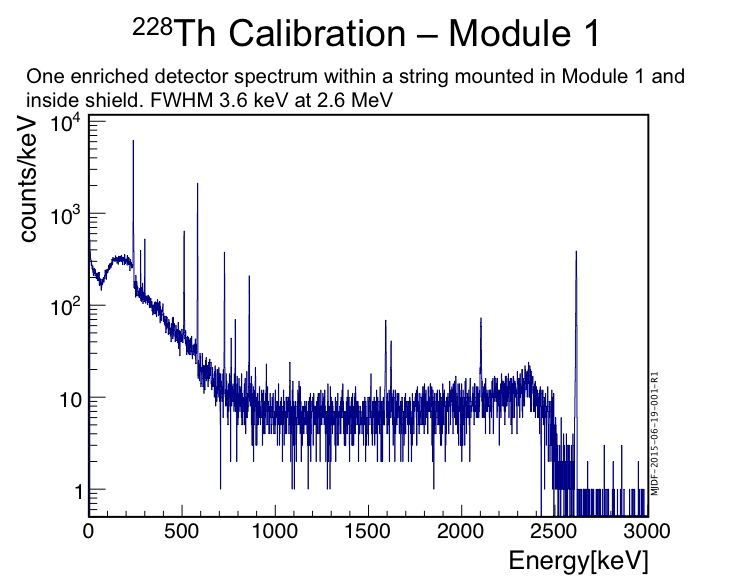}
\caption{A $^{228}$Th calibration with an enriched detector in Module~1, with incomplete shielding in place.}
\label{fig:Mod1_calib}
\end{figure}

Module~1, which included the first of two ultra-clean cryostats, was moved into the shield at the end of May 2015. It is now in commissioning with 23 of 29 detectors operating, giving 14~kg of enriched mass and 3.7~kg of natural germanium. A calibration spectrum for one of the enriched detectors in Module~1 can be seen in Fig.~\ref{fig:Mod1_calib}. The module is currently taking data without the inner electroformed copper shield in place. It will be removed from the shield and return to the glovebox once more, to allow for the installation of low-background gaskets and repair of the inoperable detectors. Blinded data-taking with Module~1 will begin shortly. 
\

Module~2, the final stage of the \textsc{Demonstrator}, is in construction; the first detector strings have been built, and the vacuum system is nearing completion. Module~2 is planned to be in commissioning by the end of 2015. 

\Acknowledgments
This material is based upon work supported by the U.S. Department of Energy, Office of Science, Office of Nuclear Physics. We acknowledge support from the Particle Astrophysics Program of the National Science Foundation. This research uses these US DOE Office of Science User Facilities: the National Energy Research Scientific Computing Center and the Oak Ridge Leadership Computing Facility. We acknowledge support from the Russian Foundation for Basic Research. We thank our hosts and colleagues at the Sanford Underground Research Facility for their support. This material is based upon work supported by the National Science Foundation Graduate Research Fellowship Program under Grant No. DGE 1256082. 
\FloatBarrier

\end{document}